\documentstyle[12pt]{article}
 at 10truept

\def\be{\beta}
\def\ga{\gamma}
\def\de{\delta}
\def\ep{\epsilon}
\def\ve{\varepsilon}

\def\et{\eta}

\def\ka{\kappa}
\def\la{\lambda}

\def\si{\sigma}

\def\ta{\tau}

\def\ph{\phi}

\def\ch{\chi}

\def\De{\Delta}

\def\Om{\Omega}

\def\fr#1#2{{{#1} \over {#2}}}

\def\bra#1{\langle{#1}|}
\def\ket#1{|{#1}\rangle}

\def\half{{\textstyle{1\over 2}}}
\def\frac#1#2{{\textstyle{{#1}\over {#2}}}}

\def\lsim{\mathrel{\rlap{\lower4pt\hbox{\hskip1pt$\sim$}}
    \raise1pt\hbox{$<$}}}
\def\gsim{\mathrel{\rlap{\lower4pt\hbox{\hskip1pt$\sim$}}
    \raise1pt\hbox{$>$}}}
\def\sqr#1#2{{\vcenter{\vbox{\hrule height.#2pt
         \hbox{\vrule width.#2pt height#1pt \kern#1pt
         \vrule width.#2pt}
         \hrule height.#2pt}}}}

\def\pr#1{{#1}^\prime}

\def\xe{ ^{129}{\rm Xe} }
\def\he{ ^{3}{\rm He} }

\def\tmf{ \widetilde{m}_F }
\def\hmf{ \widehat{m}_F }

\def\tildeb{\tilde{b}}
\def\tildec{\tilde{c}}
\def\tilded{\tilde{d}}
\def\tildeg{\tilde{g}}

\def\x{{\hat x}} 
\def\y{{\hat y}} 
\def\z{{\hat z}}

\def\X{{\hat X}}
\def\Y{{\hat Y}}
\def\Z{{\hat Z}}

\newcommand{\beq}{\begin{equation}}
\newcommand{\eeq}{\end{equation}}
\newcommand{\bea}{\begin{eqnarray}}
\newcommand{\eea}{\end{eqnarray}}
\newcommand{\rf}[1]{(\ref{#1})}

\renewenvironment{thebibliography}[1]
 { \rm
   \begin{list}{\arabic{enumi}.}
    {\usecounter{enumi} \setlength{\parsep}{0pt}
     \setlength{\itemsep}{3pt} \settowidth{\labelwidth}{#1.}
     \sloppy
    }}{\end{list}}

\begin{document}

\baselineskip=16pt
\begin{flushright}
{IUHET 420\\}
{February 2000\\}
\end{flushright}
\vglue 0.5 truein 

\begin{flushleft}
{\bf RECENT RESULTS IN LORENTZ AND CPT TESTS%
\footnote{
Presented at Orbis Scientiae 1999, 
Fort Lauderdale, Florida, December 1997}
\\}
\end{flushleft}

\vglue 0.8cm
\begin{flushleft}
{\hskip 1 truein
V. Alan Kosteleck\'y
\\}
\bigskip
{\hskip 1 truein
Physics Department\\}
{\hskip 1 truein
Indiana University\\}
{\hskip 1 truein
Bloomington, IN 47405\\}
{\hskip 1 truein
U.S.A.\\}
\end{flushleft}

\vglue 0.8cm

\noindent
{\bf INTRODUCTION}
\vglue 0.4 cm 

At a fundamental level,
nature appears invariant under Lorentz transformations.
This symmetry,
which includes rotations and boosts,
is incorporated into the standard model of particle physics.
Like other local relativistic field theories of point particles,
the standard model is also invariant under
the CPT transformation,
which is formed from the combination of charge conjugation C, 
parity reflection P, and time reversal T.
Numerous experimental tests 
of Lorentz and CPT symmetry have been performed 
\cite{pdg,cpt98}.
The exceptional sensitivity of these tests
and the cornerstone role of
Lorentz and CPT symmetry in established theory
make studies of possible Lorentz and CPT violation
of interest in the context of physics beyond the standard model
\cite{kps}.

Talks at previous conferences in this series
(Orbis Scientiae 1997-I
\cite{os1},
1997-II
\cite{os2},
and 1998)
have presented the idea that Lorentz and CPT symmetry
might be spontaneously broken in nature by effects
emerging from a fundamental theory beyond the standard model,
such as string theory
\cite{kps}.
They also have outlined the low-energy description of the resulting effects
and have described a candidate consistent standard-model extension 
incorporating Lorentz and CPT violation 
\cite{ck}.
In this talk,
I summarize some 
of the recent experimental constraints
on the standard-model extension that
that have been obtained since the previous conference.
New constraints on Lorentz and CPT violation 
are also being announced for the first time at this meeting,
as reported in other contributions to the proceedings 
\cite{bh,rw}.

Since the natual dimensionless suppression factor 
for observable Lorentz or CPT violation
is the ratio $r \sim 10^{-17}$
of the low-energy scale to the Planck scale,
relatively few experimental tests are capable of
detecting any associated effect.
Among those with the necessary sensitivity 
and placing interesting constraints on parameters
in the standard-model extension are 
studies of neutral-meson oscillations 
\cite{kexpt,kp,ckpv,bexpt,ak,ktevproc},
comparative tests of QED 
in Penning traps \cite{pennexpts,bkr,gg,hd,rm},
spectroscopy of hydrogen and antihydrogen \cite{bkr2,rw,rbproc},
measurements of muon properties
\cite{bkl},
clock-comparison experiments
\cite{ccexpt,kl,rw},
observations of the behavior of a spin-polarized torsion pendulum
\cite{bk,bh},
measurements of cosmological birefringence \cite{cfj,ck,jk,pvc},
and observations of the baryon asymmetry \cite{bckp}.
Effects on cosmic rays have also been investigated 
in a restricted version of the standard-model extension
\cite{cg,be}.
In this contribution to the proceedings,
I limit considerations to
recent results directly relevant to the standard-model extension 
and obtained in kaon oscillations
and in clock-comparison experiments.

\vglue 0.6 cm 
\noindent
{\bf EXPERIMENTS WITH NEUTRAL KAONS}
\vglue 0.4 cm 

Neutral-meson oscillations provide a sensitive tool
for studies of Lorentz and CPT symmetry.
In the kaon system,
experiments already constrain
the CPT figure of merit
$r_K \equiv |m_K - m_{\overline{K}}|/m_K$
to better than a part in $10^{18}$
\cite{kexpt,ktevproc,cplearproc},
with improvements expected in the near future
\cite{kloe}.

The standard analysis 
\cite{lw,rs}
of possible CPT violation
in the kaon system is purely phenomenological,
introducing a complex parameter $\de_K$
in the standard relationships between the physical meson states
and the strong-interaction eigenstates.
However,
in the context of the standard-model extension
with Lorentz and CPT violation,
the parameter $\de_K$
is calculable rather than purely phenomenological.
Thus,
a meson with velocity $\vec\be$ in the laboratory frame
and associated boost factor $\ga$
displays CPT-violating effects given by
\cite{ak}
\beq
\de_K \approx i \sin\hat\ph ~ e^{i\hat\ph} 
\ga(\De a_0 - \vec \be \cdot \De \vec a) /\De m
\quad .
\label{dek}
\eeq
In this expression,
$\hat\ph \equiv \tan^{-1}(2\De m/\De \ga )$,
where $\De m$ and $\De\ga$ are the mass and rate differences
between the physical eigenstates,
and the four components of the quantity $\De a_\mu$ 
control certain specific
Lorentz- and CPT-violating couplings in the standard-model extension.

Equation \rf{dek}
exhibits several unexpected features,
including dependence on momentum magnitude and orientation.
These imply various observable consequences including,
for example,
time variations of the measured value of $\de_K$ 
with the Earth's sidereal (not solar) rotation frequency 
$\Om \simeq{2\pi}$/(23 h 56 min)
\cite{ak}.
To display explicitly the time dependence of $\de_K$ 
arising from the rotation of the Earth,
one can introduce a convenient nonrotating frame.
Denote the nonrotating-frame basis 
as $(\X,\Y,\Z)$.
The natural choice for $\Z$ is the rotation axis of the Earth,
and celestial equatorial coordinates
can be used to fix the $\X$ and $\Y$ axes
\cite{kl}.

For the general case of a kaon with three-velocity $\vec \be$
in the laboratory frame,
an expression for the parameter $\de_K$ in the nonrotating 
frame can be found.
However,
for simplicity in what follows
I restrict attention to the special case of
experiments involving highly collimated uncorrelated kaons 
with nontrivial momentum spectrum and large mean boost,
such as the E773 and KTeV experiments
\cite{kexpt,ktev}
where the average boost factor $\overline\ga$ is of order 100.
General theoretical expressions and a discussion of
issues relevant to other classes of experiment
can be found in Ref.\ \cite{ak}.

In experiments with boosted collimated kaons,
the $\z$ axis for the laboratory frame 
can be chosen along the kaon three-velocity,
$\vec\be = (0,0,\be )$.
The general expression for $\de_K$ in the nonrotating frame 
then reduces to 
\beq
\de_K (\vec p, t) = 
\fr {i \sin\hat\ph ~ e^{i\hat\ph}} {\De m} \ga
[ \De a_0 + \be \De a_Z \cos\ch 
+ \be \sin\ch ( \De a_Y \sin\Om t + \De a_X \cos\Om t ) ],
\label{deptktev}
\eeq
where $\cos{\ch}=\z\cdot\Z$.
Note that this equation shows that 
each component of $\De a_\mu$ in the nonrotating frame
is associated with momentum dependence through the boost factor $\ga$,
but only the coefficients of $\De a_X$ and $\De a_Y$
vary with sidereal time. 

Experiments are performed over extended time periods,
so a conventional analysis for CPT bounds 
disregarding the momentum and time dependence
is sensitive to a time and momentum average 
over the data momentum spectrum given by 
\beq
|\overline {\de_K}| = 
\fr {\sin\hat\ph } {\De m} \overline{\ga}
( \De a_0 + \overline{\be} \De a_Z \cos\ch )
\quad ,
\label{deptktevav2}
\eeq
where 
$\overline{\be}$ and $\overline{\ga}$ 
are averages of $\be$ and $\ga$.
This expression allows the derivation of a bound
on a combination of the quantities
$\De a_0$ and $\De a_Z$
\cite{ak}:
\beq
|\De a_0 + 0.6 \De a_Z| \lsim 10^{-20} {\rm ~ GeV}
\quad .
\label{bound1}
\eeq 

In practice,
the experimental constraints on $\de_K$ 
are obtained via measurements on other observables
such as the mass difference $\De m$,
the $K_S$ lifetime $\ta_S= 1/\ga_S$,
and the ratios $\et_{+-}$, $\et_{00}$ 
of amplitudes for $2\pi$ decays.
Analysis shows that 
only the phases $\ph_{+-}$ and $\ph_{00}$ of the latter 
vary with momentum and sidereal time at leading order
\cite{ak}.
For example,
the phase $\ph_{+-}$ is given by
\beq
\ph_{+-} \approx \hat\ph + 
\fr {\sin\hat\ph } {|\et_{+-}| \De m} \ga
[ \De a_0 + \be \De a_Z \cos\ch 
+ \be \sin\ch ( \De a_Y \sin\Om t + \De a_X \cos\Om t) ] .
\label{deptktev3}
\eeq
This expression shows that
distinct bounds on the components of $\De a_\mu$
can in principle be obtained
in experiments with boosted collimated kaons 
if the momentum spectrum is sufficiently resolved.

The KTeV collaboration has recently placed a 
constraint $A_{+-}\lsim 0.5^\circ$
on the amplitude $A_{+-}$ of time variations of the phase $\ph_{+-}$
with sidereal periodicity 
\cite{ktevproc}.
This gives the limit 
\beq
\sqrt{(\De a_X)^2 + (\De a_Y)^2} \lsim 10^{-20} {\rm ~GeV}
\quad ,
\label{bound2}
\eeq
which represents the first bound obtained 
on the parameters $\De a_X$ and $\De a_Y$.

It should be noted that experiments with neutral mesons 
are presently the only ones 
known to be capable of detecting effects
associated with the Lorentz- and CPT-violating parameter $\De a_\mu$
\cite{ak}.
Note also that the two bounds \rf{bound1} and \rf{bound2} discussed here
are independent constraints on possible CPT violation.
Relative to the kaon mass,
both bounds compare favorably with the ratio of the kaon mass to
the Planck scale.

\vglue 0.6 cm
\noindent
{\bf CLOCK-COMPARISON EXPERIMENTS}
\vglue 0.4 cm

Among the most sensitive tests of Lorentz and CPT symmetry are 
the clock-comparison experiments
\cite{ccexpt}.
These provide limits on possible spatial anisotropies
and hence on violations of rotation symmetry
by bounding the relative frequency change
between two hyperfine or Zeeman transitions
as the Earth rotates.
Data from these experiments can be interpreted
in the context of the standard-model extension
\cite{kl}.
In this section,
I provide a brief outline of the primary results of this study. 

The standard-model extension allows for flavor-dependent effects.
Since distinct species of atoms and ions 
have different compositions in terms of elementary particles,
the corresponding signals in clock-comparison experiments
can crucially depend on the chosen species. 
The complexity of most atoms and ions makes it impractical
to perform a complete \it ab initio \rm
calculation of energy-level shifts 
arising from Lorentz-violating terms in the standard-model extension.
Fortunately,
since any Lorentz-violating effects must be minuscule,
it suffices to determine leading-order effects 
in a perturbative calculation.
The leading perturbative contribution 
to Lorentz-violating energy-level shifts 
consists of a sum of shifts originating 
from each elementary particle in the atom,
generated through the expectation value
of the nonrelativistic Lorentz-violating hamiltonian 
in the multiparticle unperturbed atomic state. 

The appropriate single-particle nonrelativistic hamiltonian 
is known \cite{fwpaper},
and its perturbation component $\de h$ for Lorentz violation
has the form
\beq
\de h 
= \left( a_0 -m c_{00} -m e_0 \right) 
+\left( -b_j + m d_{j0} - \half m \ve_{jkl}g_{kl0} 
      + \half \ve_{jkl}H_{kl} \right) \si^j  
+\ldots
   \quad .
\label{nrham3}
\eeq 
Here,
$m$ is the single-particle mass,
each Lorentz index is split into a timelike component 0
and spacelike cartesian components $j = 1,2,3$,
$\ve_{jkl}$ is the totally antisymmetric rotation tensor,
and the Pauli matrices are denoted by $\si^j$.
The other quantities are parameters for Lorentz and CPT violation
arising in the standard-model extension.
A complete expression for $\de h$ is given
in Refs.\ \cite{kl,fwpaper}.

The multiparticle Lorentz-violating perturbative hamiltonian 
describing an atom $W$
is formed as the sum of the perturbative hamiltonians
for each of the $N_w$ particles of type $w$ comprising $W$:
\beq
\pr{h}=\sum_w\sum_{N=1}^{N_w} \de {h}_{w,N}
\quad .
\label{hprime}
\eeq
Here, $w$ is $p$ for the proton,
$n$ for the neutron, and $e$ for the electron.
The perturbative hamiltonian $\de {h}_{w,N}$ 
for the $N$th particle of type $w$
has the form given in Eq.\ \rf{nrham3},
except that the dependence 
of the parameters for Lorentz violation
is shown by a superscript $w$.

The perturbative Lorentz-violating energy shift of the state $\ket{F,m_F}$
of $W$ is derived as the expectation value
$\bra{F, m_F}\pr{h}\ket{F, m_F}$
of the perturbative hamiltonian \rf{hprime} in the appropriate 
unperturbed quantum state.
After some calculation,
one finds \cite{kl}
\beq 
\bra{F,m_F} \pr{h} \ket{F,m_F} = 
\hmf E_d^W + \tmf E_q^W 
\quad .
\label{shift}
\eeq
Here,
$\hmf$ and $\tmf$
are ratios of Clebsch-Gordan coefficients
\cite{kl}.
The dipole and quadrupole energy shifts
$E_d^W$ and $E_q^W$ 
are explicitly given in Ref.\ \cite{kl}, 
and they involve
components of the parameters for Lorentz violation
defined in the laboratory frame.

Since the laboratory frame rotates with the Earth,
the laboratory-frame components change cyclically 
with the Earth's sidereal rotation frequency 
$\Om$.
It is therefore more convenient to work in a nonrotating frame.
Denote the nonrotating-frame basis 
by $(\X,\Y,\Z)$ as in the previous section,
and let the laboratory-frame basis be
$(\x,\y,\z)$.
The $\z$ axis here
is taken as the quantization axis of $W$ for the given experiment,
and the angle $\ch\in(0,\pi)$
given by $\cos{\ch}=\z\cdot\Z$ is assumed nonzero.

To express the results in a relatively compact form,
it is convenient to introduce nonrotating-frame 
combinations of Lorentz-violating parameters,
denoted
$\tildeb_J$,
$\tildec_{Q}$,
$\tildec_{Q,J}$,
$\tildec_{-}$,
$\tildec_{XY}$,
$\tilded_J$,
$\tildeg_{D,J}$,
$\tildeg_{Q}$,
$\tildeg_{Q,J}$,
$\tildeg_{-}$,
$\tildeg_{XY}$.
Their definitions in terms of quantities in
the nonrelativistic hamiltonian $h$
can be found in Ref.\ \cite{kl}.
As one example, 
\beq
\tildeb^w_J := b^w_J -m d^w_{J0} 
 +\half m\ep_{JKL} g^w_{KL0} -\half \ep_{JKL} H^w_{KL}
 \quad ,
\label{bdcgtildenonrot}
\eeq
which involves a combination 
of CPT-odd and CPT-even couplings in the standard-model extension.
Here,
spatial indices in the nonrotating frame 
are denoted by $J = X, Y, Z$,
the time index is denoted $0$,
and $\ep_{JKL}$ is the nonrotating-frame antisymmetric tensor. 

Substituting the above into the expression 
for the energy-level shift gives
\beq
\bra{F,m_F} \pr{h} \ket{F,m_F} = E_0
+E_{1X}\cos{\Om t}+E_{1Y}\sin{\Om t}
 + E_{2X}\cos{2\Om t}+E_{2Y}\sin{2\Om t}
\quad .
\label{tdatoms}
\eeq
The energy $E_0$ is constant in time 
and is therefore irrelevant for clock-comparison experiments.
The four other energies are given explicitly 
in terms of the Lorentz-violating parameters
and other quantities in Ref.\ \cite{kl}. 
In clock-comparison experiments,
the result is typically  
a bound on the
amplitude of the time variation of a transition frequency,
determined here as the difference between two energy-level shifts 
of the form $\bra{F,m_F} \pr{h} \ket{F,m_F}$.

In the remainder of this section,
I consider the clock-comparison experiments
performed by
Prestage {\it et al.},
Lamoreaux {\it et al.},
Chupp {\it et al.},
and Berglund {\it et al.} \cite{ccexpt}.
Each of the bounds from each of these experiments
fits one of the following forms:
\bea
&& 
 \Big| 
 \sum_{w} 
  [ u_0^A( \be_w^A \tildeb_X^w + \de_w^A \tilded_X^w 
        + \ka_w^A \tildeg_{D,X}^w )
   + u_1^A( \ga_w^A \tildec_{Q,X}^w 
           + \la_w^A \tildeg_{Q,X}^w ) ]
 \nonumber\\
  &&
  - v\sum_{w} 
    [ u_0^B( \be_w^B \tildeb_X^w + \de_w^B \tilded_X^w 
           + \ka_w^B \tildeg_{D,X}^w )
     + u_1^B( \ga_w^B \tildec_{Q,X}^w 
           + \la_w^B \tildeg_{Q,X}^w ) ]
  \Big| 
\lsim 2\pi\ve_{1,X} ,
   \nonumber \\
&& 
 \Big| 
 \sum_{w} 
  [ u_0^A( \be_w^A \tildeb_Y^w + \de_w^A \tilded_Y^w 
        + \ka_w^A \tildeg_{D,Y}^w )
   + u_1^A( \ga_w^A \tildec_{Q,Y}^w 
           + \la_w^A \tildeg_{Q,Y}^w ) ]
 \nonumber\\
  &&
  - v\sum_{w} 
    [ u_0^B( \be_w^B \tildeb_Y^w + \de_w^B \tilded_Y^w 
           + \ka_w^B \tildeg_{D,Y}^w )
     + u_1^B( \ga_w^B \tildec_{Q,Y}^w 
           + \la_w^B \tildeg_{Q,Y}^w ) ]
  \Big| 
\lsim 2\pi\ve_{1,Y} , 
   \nonumber \\
&& 
 \Big| \sum_{w} 
  u_2^A
  ( \ga_w^A \tildec_-^w 
   +\la_w^A \tildeg_-^w )
  -v \sum_{w}
  u_2^B
  ( \ga_w^B \tildec_-^w 
   +\la_w^B \tildeg_-^w )
 \Big|
\lsim 2\pi \ve_{2,-} ,
   \nonumber \\
&&
 \Big| \sum_{w} 
  u_2^A
  ( \ga_w^A \tildec_{XY}^w
   +\la_w^A \tildeg_{XY}^w )
  -v \sum_{w}
  u_2^B
  ( \ga_w^B \tildec_{XY}^w 
   +\la_w^B \tildeg_{XY}^w )
 \Big| 
\lsim 2\pi \ve_{2,XY} .
\label{generalbound}
\eea
In these expressions,
the coefficients $u_0$, $u_1$, $u_2$, and $v$
contain the dependences on quantities such as 
$\hmf$, $\tmf$, $\ch$, and gyromagnetic ratios.
The quantities $\be$, $\de$, $\ka$, $\ga$, $\la$
with superscripts and subscripts are special matrix elements
described in Ref.\ \cite{kl}.
The parameter $v=g_A/g_B$ is the ratio 
of gyromagnetic ratios for the atomic species $A$ and $B$
involved in the particular experiment.
The associated bounds on the amplitudes 
of frequency shifts are denoted 
$\ve_{1,X}$,
$\ve_{1,Y}$,
$\ve_{2,-}$, 
$\ve_{2,XY}$, 
corresponding to sidereal or semi-sidereal variations as 
$\cos \Om t$,
$\sin \Om t$, 
$\cos 2\Om t$,
$\sin 2\Om t$, 
respectively.
All the quantities in the above experiment
are tabulated in Ref.\ \cite{kl}
for each of the experiments in question.

It turns out that the experimental results
all constrain distinct
linear combinations of parameters for Lorentz violation.
A useful tool for studying 
specific sensitivities is
the nuclear Schmidt model
\cite{schmidt}.
In this context,
the Prestage {\it et al.},
Lamoreaux {\it et al.},
and Chupp {\it et al.}
experiments are sensitive 
to neutron parameters for Lorentz violation,
while the Berglund {\it et al.} experiment  
is sensitive to electron, proton, and neutron parameters.
In fact,
only a subset of the allowed parameter space is constrained
\cite{kl}
by all these experiments.
However,
the bounds obtained are impressive
and represent sensitivity to Planck-scale physics.

The relatively complicated form 
of the results \rf{generalbound} can be simplified
under certain assumptions. 
If one supposes 
both no appreciable cancellation of effects
between the species $A$ and $B$
and no cancellations among different terms in the sums
appearing in Eq.\ \rf{generalbound},
then the numerical value of each bound
can be applied to each term in the sum,
producing individual constraints on 
the parameters for Lorentz violation
appearing in Eq.\ \rf{generalbound}.
To obtain specific values,
one can work within 
the context of the Schmidt model
and make some crude dimensional estimates
of the unknown matrix elements.
The results of this procedure
are tabulated in Ref.\ \cite{kl}.
For example,
one finds that the 
Lorentz- and CPT-violating parameters $\tildeb_J^w$
are most tightly constrained by the experiment of 
Berglund {\it et al.},
which gives
$|\tildeb_J^n| \lsim 10^{-30}$ GeV,
$|\tildeb_J^e| \lsim 10^{-27}$ GeV,
$|\tildeb_J^p| \lsim 10^{-27}$ GeV.
The experiments in Ref.\ \cite{ccexpt}
also bound other parameters,
as described in Ref.\ \cite{kl}. 

Experiments producing both calculable and clean bounds
would evidently be of particular theoretical interest.
One possibility for improving both calculability and cleanliness 
is to use species $W$ for which the Lorentz-violating energy shifts depend
predominantly on a single valence particle $w$.
For example, 
in the case where $w$ is an electron,
substances of nuclear spin zero could be used. 
For the case where $w$ is a nucleon,
a list of nuclei theoretically expected
to yield relatively calculable and clean bounds
is provided in Ref.\ \cite{kl}.

\vglue 0.6 cm
\noindent
{\bf NEW RESULTS REPORTED AT THIS CONFERENCE}
\vglue 0.4 cm

In other presentations to this conference
\cite{bh,rw},
new experimental results are reported
that provide relatively calculable and clean
bounds on certain Lorentz-violating parameters
in the standard-model extension.
In this final section,
I provide a brief summary placing these results
in the context of the preceding discussion.

\it Neutron parameters. \rm
An interesting limit on neutron parameters for Lorentz violation
is attainable using a dual nuclear Zeeman $\he$-$^{129}$Xe maser 
\cite{stoner}
because the $I=\half$ nucleus $^{129}$Xe is sensitive
to dipole energy shifts from neutron parameters.
Within the Schmidt model,
the description of the $\he$ and $\xe$ systems are related,
which leads to a relatively clean bound
\cite{kl}.
At this conference,
Walsworth discusses 
\cite{rw}
an experiment producing a bound of 80 nHz on sidereal variations 
of the free-running $\he$ frequency using $\xe$ as a reference.
In the context of the Schmidt model
and the assumptions described in the previous section,
this can be interpreted as a bound on equatorial components
of $|\tildeb_J^n|$ of approximately $10^{-31}$ GeV
\cite{rw}.

\it Electron parameters. \rm
High-sensitivity tests of Lorentz symmetry in the electron sector
can be performed by searching for Lorentz-violating spin couplings
with macroscopic materials having a net spin polarization
generated by the effects of many electrons
\cite{bk}.
The most sensitive apparatus of this type at present is
the spin-polarized torsion pendulum used 
with the E\"ot-Wash II instrument
at the University of Washington
\cite{ea,harr,bh},
which involves stacked layers of toroidal magnets
producing a large net electron spin but negligible magnetic moment.
At this conference,
Heckel describes
\cite{bh}
an analysis of data taken with this apparatus
that places a strong constraint on the 
components $|\tildeb_J^e|$,
at the level of about $10^{-29}$ GeV for the equatorial components
and about $10^{-28}$ GeV for the component along $\Z$.

\it Proton parameters. \rm
Since hydrogen is theoretically well understood,
it is a good candidate for a substance
producing a calculable bound in a clock-comparison experiment.
In fact,
the reference transition in the Prestage {\it et al.} experiment
was a hydrogen maser.
In the context of the standard-model extension,
analyses of experiments with hydrogen and antihydrogen
have been performed \cite{bkr,bkr2,rbproc}.
The standard H-maser line 
involves atomic states with $m_F=0$
and is insensitive to Lorentz violation,
but the other ground-state hyperfine lines 
involve states with $m_F=\pm 1$
and therefore are sensitive to Lorentz violation.
The sidereal variations of these lines
are determined at leading order by
the combinations $\tildeb^e_J \pm \tildeb^p_J$.
At this conference,
Walsworth describes 
\cite{rw}
an experiment with hydrogen masers that places a bound 
of 0.7 mHz on the magnitude
of sidereal variations in these frequencies.
Combined with the above constraints on $\tildeb^e_J$ 
in the electron sector announced by Heckel \cite{bh},
this can be interpreted as a bound on
the equatorial components of $|\tildeb_J^p|$ 
of approximately $4 \times 10^{-27}$ GeV
\cite{rw}.

\vglue 0.6 cm
\noindent
{\bf ACKNOWLEDGMENTS}
\vglue 0.4 cm

This work is supported in part
by the United States Department of Energy 
under grant number DE-FG02-91ER40661.

\vglue 0.6 cm
\noindent
{\bf REFERENCES}
\vglue 0.4 cm

\end{document}